\documentclass[10pt,paper,twocolumn]{ieeetran}
\usepackage{epsf, epsfig}
\usepackage{amsthm}
\usepackage{latexsym}
\usepackage{amsmath,bm}

\usepackage{cases}
\usepackage{url}
\usepackage{graphicx}
\usepackage{multirow}
\usepackage{color}
\usepackage{amssymb}
\usepackage{float}
\usepackage{algorithm}
\usepackage{algorithmic}
\usepackage{cite}
\usepackage{amsthm}
\usepackage{latexsym}
\usepackage{amsmath}
\usepackage[colorlinks,linkcolor=black,anchorcolor=black,citecolor=black,urlcolor=black]{hyperref}
\hypersetup{
bookmarksnumbered=true,
pdfstartview=FitH
}
\newcommand\normt[1]{\left\lVert#1\right\rVert_2}
\newcommand\normf[1]{\left\lVert#1\right\rVert_F^2}
\newcommand\norma[1]{\left\lVert#1\right\rVert_{\mathcal{A}}}

\begin{document}

\title{\huge{Super-Resolution Channel Estimation for Arbitrary Arrays in Hybrid Millimeter-Wave Massive MIMO Systems}}

\author{Yue~Wang,~\IEEEmembership{Member,~IEEE}, Yu~Zhang, Zhi~Tian,~\IEEEmembership{Fellow,~IEEE}, Geert~Leus,~\IEEEmembership{Fellow,~IEEE}, and Gong~Zhang
\IEEEcompsocitemizethanks{
\IEEEcompsocthanksitem This work was supported in part by the US National Science Foundation (NSF) grants \#1547364, \#1527396, \#1546604, and \#1730083, and the National Science Foundation of China (NSFC) grant \#61871218. This work was partly carried out in the frame of the ASPIRE project (project 14926 within the OTP program of NWO-TTW).
}
\IEEEcompsocitemizethanks{
\IEEEcompsocthanksitem Y. Wang and Z. Tian are with the Department of Electrical and Computer Engineering, George Mason University, Fairfax, VA 22030, USA. (e-mail: ywang56@gmu.edu; ztian1@gmu.edu)
}
\IEEEcompsocitemizethanks{
\IEEEcompsocthanksitem Y. Zhang and G. Zhang are with the Key Lab of Radar Imaging and Microwave Photonics, Ministry of Education, Nanjing University of Aeronautics and Astronautics, Nanjing 211100, China. (e-mail: skywalker\_zy@163.com; gzhang@nuaa.edu.cn)
}
\IEEEcompsocitemizethanks{
\IEEEcompsocthanksitem G. Leus is with the Faculty of Electrical Engineering, Mathematics and Computer Science, Delft University of Technology, Delft 2826 CD, The Netherlands. (e-mail: g.j.t.leus@tudelft.nl)
}
}

\maketitle

\begin{abstract}
This paper develops efficient channel estimation techniques for millimeter-wave (mmWave) massive multiple-input multiple-output (MIMO) systems under practical hardware limitations, including an arbitrary array geometry and a hybrid hardware structure. Taking on an angle-based approach, this work adopts a generalized array manifold separation approach via Jacobi-Anger approximation, which transforms a non-ideal, non-uniform array manifold into a virtual array domain with a desired uniform geometric structure to facilitate super-resolution angle estimation and channel acquisition. Accordingly, structure-based optimization techniques are developed to effectively estimate both the channel covariance and the instantaneous channel state information (CSI) within a short sensing time. In particular, the difference in time-variation of channel path angles and path gains is capitalized to design a two-step CSI estimation scheme that can quickly sense fading channels. Theoretical results are provided on the fundamental limits of the proposed technique in terms of sample efficiency. For computational efficiency, a fast iterative algorithm is developed via the alternating direction method of multipliers. Other related issues such as spurious-peak cancellation in nonuniform linear arrays and extensions to higher-dimensional cases are also discussed. Simulations testify the effectiveness of the proposed approaches in hybrid mmWave massive MIMO systems with arbitrary arrays.
\end{abstract}

\section{Introduction}
In millimeter-wave (mmWave) massive multiple-input multiple-output (MIMO) communications, large antenna gains coupled with the availability of large bandwidths bring many desired benefits such as high throughput, large capacity, and robustness against fading and interference \cite{Larsson2014Massive, Sun2014MIMO}, which all hinge on accurate channel knowledge. However, the 
increase in antennas 
results in an enlarged channel dimension that gives rise to 
challenges to traditional channel estimation techniques, in terms of the high signal acquisition cost and the large training overhead
\cite{Rusek2013Scaling, Rangan2014Millimeter}.
The hinderance in channel estimation is further aggravated by practical hardware limitations.
A hybrid analog-digital architecture is widely suggested for massive MIMO transceivers, which reduces the number of radio frequency (RF) chains by balancing between the analog RF part and the digital baseband part \cite{Molisch2017Hybrid, Alkhateeb2014MIMO}. However, under such a hybrid structure, the channel estimator at baseband can only observe a compressed representation of the channel through a few RF chains.

To overcome these challenges, compressive sensing (CS) has been advocated for channel estimation in mmWave massive MIMO systems \cite{Bajwa2010Compressed, Schniter2014Channel, Alkhateeb2014Channel, Lee2016Channel, Mendez-Rial2015Channel, Gao2016Channel, Wang2016Fast}. These CS-based approaches exploit the channel sparsity that stems from the limited scattering characteristics of mmWave propagation \cite{Shafi2018Microwave, Rappaport2014Millimeter, Zhang2010Channel, Rappaport2013Broadband}. Through virtual channel modeling \cite{Sayeed2002Deconstructing}, the large-dimensional mmWave massive MIMO channels can be represented by only a small number of parameters, including the angles of departure/arrival (AoD/AoA) and the path gains of the sparse scattering paths. Therefore, CS techniques enable channel estimation from a small set of compressively collected training samples. In \cite{Bajwa2010Compressed, Schniter2014Channel}, a sparse multipath channel is formulated as a sparse vector on the angle-delay-Doppler space, and then CS techniques are applied to recover the vectorized sparse channel. In \cite{Alkhateeb2014Channel}, an adaptive CS-based algorithm is proposed to estimate the sparse channel with a hybrid analog-digital hardware architecture. In \cite{Lee2016Channel}, a hybrid architecture based on phase shifters is proposed to recover the sparse channel via greedy search algorithms. To further reduce 
the power consumption of phase shifters, a switch-based hybrid architecture is developed in \cite{Mendez-Rial2015Channel}, for sparse channel estimation. In \cite{Gao2016Channel}, the CS-based channel estimation scheme is extended to broadband mmWave MIMO systems. In \cite{Wang2016Fast}, to reduce the problem complexity, CS-based channel estimation is divided into angle estimation and path gain estimation subproblems, which are solved sequentially.

All the aforementioned techniques aim to estimate the instantaneous channel state information (CSI). Another line of work focuses on estimating the 
channel statistics, such as the channel covariance \cite{Wang2016Efficient, Park2018Spatial}. The channel covariance is an important second-order statistic, which remains constant over many channel coherence intervals and therefore can be used for statistics-based design of the precoders, beamformers and linear receivers \cite{Park2017Exploiting, Li2017Optimizing}. To estimate the second-order statistics of the vectorized sparse mmWave MIMO channel, a diagonal-search orthogonal matching pursuit algorithm is developed in \cite{Wang2016Efficient}, which not only utilizes the joint sparsity represented by the available multiple measurement vectors (MMV) but also takes advantage of the Hermitian structure of the channel covariance matrix. In \cite{Park2018Spatial}, a CS-based channel covariance estimator is proposed by using dynamic sensing schemes and designing dynamic greedy pursuit algorithms for the hybrid architecture.

Existing CS-based channel estimators critically hinge on an on-grid assumption that the values of the AoD/AoA of each propagation path exactly reside on some predefined grid in the angular domain. However, in practice, the AoD/AoA of paths are continuously-valued off grid. As a result, CS-based methods suffer from degraded performance due to the power leakage effect around the recovered discrete grid points, a.k.a., the infamous basis mismatch problem \cite{Chi2011Sensitivity}. An angle rotation technique is proposed to alleviate this problem, which is developed upon the spatial basis expansion model \cite{Xie2016UL/DL, Xie2017Unified}. It improves the estimation accuracy in the angular domain, but still experiences finite resolution due to some predefined spatial rotation parameters.
A continuous basis pursuit technique is proposed for perturbed CS in \cite{Zhu2011Sparsity}, which is however limited through the series expansion.
On the other hand, classical subspace methods, such as MUSIC and ESPRIT, can achieve super-resolution in angle estimation \cite{Schmidt1986Multiple, Roy1989ESPRIT}. But, they require a large number of snapshots for collecting sample statistics, which leads to a long sensing time and consumes large training resources. To circumvent the on-grid assumption required by traditional CS and achieve super-resolution at short sensing time, a gridless CS technique is developed via atomic norm minimization (ANM) in the form of semidefinite programming (SDP) \cite{Chandrasekaran2012Convex, Candes2014Towards, Tang2013Compressed}. As a structure-based optimization technique, gridless CS is applied for super-resolution channel estimation in mmWave massive MIMO systems \cite{Wang2017Efficient, Tsai2018Millimeter, Haghighatshoar2017Massive}, which utilizes not only the sparsity of the channels but also the Vandermonde structure of the antenna arrays.

By capitalizing on the critical Vandermonde structure, gridless CS implicitly assumes the use of an ideal uniform array geometry, that is, the antennas have to be uniformly placed with exactly the same separation distance. However, in practical applications, arbitrary arrays arise in several cases, instead of the perfect uniform arrays. For example, the antenna separation distance is measured in the millimeter range over the mmWave frequency bands. Thus, an 
ideal uniform array geometry is hard to guarantee due to calibration errors introduced in the manufacturing process and/or antenna installation. Another case of arbitrary arrays appears due to sub-array selection. For example, for the purpose of energy saving in the switch-based hybrid architecture, only a small number of antennas is switched to link the RF chains \cite{Mendez-Rial2016Hybrid}. Arbitrary arrays no longer present the well-featured Vandermonde structure explicitly in the array manifold, which then excludes the use of a large number of geometric-based channel estimation techniques.
To overcome this problem, array manifold separation techniques have been developed in array signal processing \cite{Belloni2007DoA, Gershman2010One, GovindaRaj2019Single}. However, the structural feature presented via manifold separation is not efficiently utilized in \cite{Belloni2007DoA, Gershman2010One}, where conventional subspace methods are used based on large samples. In \cite{GovindaRaj2019Single}, since the manifold separation through the Fourier series approximation is employed to enable ANM, it suffers from an expensive computational complexity in order to reduce the approximation error in the largely 
expanded Fourier domain.

In this paper, addressing all the aforementioned challenges cohesively, we seek to design high-performance, low-cost channel estimation solutions for arbitrary arrays in hybrid mmWave massive MIMO systems. Specifically, we propose a super-resolution channel estimation framework that not only utilizes the special channel features of the sparse mmWave massive MIMO propagation, but also fully considers the nonideal array geometry and practical hardware limitations. This framework offers several channel estimation solutions and enables to obtain both the channel statistics and the instantaneous CSI, depending on whether the transceiver design is built on channel covariance \cite{Park2017Exploiting, Li2017Optimizing} or the channel itself \cite{ Ayach2014Spatially}. We propose two super-resolution solutions for channel covariance estimation (CCE) through efficient structure-based optimization techniques, with samples collected from multiple snapshots. One is the CCE via the low-rank structured covariance reconstruction (LRSCR), which provides super-resolution accuracy at a low computational cost. The other is the CCE via Dynamic-ANM, which further allows for a dynamic configuration where the hybrid hardware parameters change over time for a better performance.
For block fading channels, given the estimate from CCE, the instantaneous CSI can then be estimated in a timely fashion.
Since the angles change slowly and can take a long time to acquire accurately from the channel statistics while the path gains vary frequently but are easy to acquire given the estimated angles, the instantaneous CSI estimation is divided by solving two subproblems sequentially, i.e., angle estimation and path gain estimation.
In developing these novel super-resolution channel estimation approaches, this work contains the following main contributions.
\vspace{-0.01in}
\begin{itemize}
\item We leverage a generalized array manifold separation approach to extract the useful geometric structure for a practical system with an arbitrary or imperfect array geometry. In particular, we transform the sparse mmWave massive MIMO channel representation from the physical arbitrary antenna domain to a virtual uniform antenna domain via the Jacobi-Anger approximation \cite{Abramowitz1964Handbook}. Our method enables gridless CS to exploit the useful Vandermonde structure presented in the virtual uniform array manifold.

\item 
    This work not only develops super-resolution channel estimation solutions, but also investigates the fundamental limits of gridless CS based channel estimation under 
    the constraints of arbitrary arrays and hybrid structures. Our theoretical results shed light on 
    the minimum number of RF chains required by super-resolution channel estimation, as well as the lower and upper bounds on the mode order selected for the Jacobi-Anger approximation.
    This leads to a tradeoff between the hardware cost of sparse channel estimation and the approximation accuracy to combat the imperfect array geometry.

\item To reduce the high computational complexity of the SDP-based channel estimation, we design a fast iterative algorithm through the alternating direction method of multipliers (ADMM) \cite{Boyd2010Distributed}. It provides an efficient first-order algorithm implementation with much lower computational complexity than that of the SDP solvers using the interior-point method.

\item We tackle several practical implementation issues. 
    Specifically, we overcome the side effect of the array manifold separation operation, by removing the spurious peaks generated by the Jacobi-Anger approximation for nonuniform linear arrays. We also extend our work to more complicated 2-dimensional (2D) scenarios, where both the BS and the MS are equipped with multiple antennas.
\end{itemize}

Simulation results are provided to testify the advantages of the proposed solutions, which make them attractive for hybrid mmWave massive MIMO systems with arbitrary arrays.

The rest of this paper is organized as follows. Section~\ref{sect:modelpriorart} presents the system model and problem formulation for sparse channel estimation in hybrid arbitrary arrays. Section~\ref{sect:superChEst} proposes a super-resolution channel estimation framework based on the array manifold separation, in which different channel estimation solutions are developed for obtaining the channel covariance and the instantaneous CSI. Specific issues related to the proposed techniques are discussed in Section~\ref{sect:discussions}. Simulation results are presented in Section~\ref{sect:simulations}, followed by conclusions in Section~\ref{sect:conclusions}.

\textit{Notations:} $a$ is a scalar, $\bm{a}$ denotes a vector, $\bm{A}$ is a matrix, and $\mathcal{A}$ represents a set. $(\cdot)^T$, $(\cdot)^*$, and $(\cdot)^H$ are the transpose, conjugate, and conjugate transpose of a vector or matrix, respectively. $\text{conv}(\mathcal{A})$ means the convex hull of a set $\mathcal{A}$. $\text{Real}(\cdot)$ and $\text{Imag}(\cdot)$ compute the real part and the imaginary part of a vector or matrix, respectively. $|a|$ denotes the absolute value of $a$. $\normt{\bm{a}}$ is the $\ell_2$ norm of $\bm{a}$. $\text{diag}(\bm{a})$ and $\text{diag}(\bm{A}_1,\bm{A}_2)$ denote a diagonal matrix with the diagonal elements constructed from $\bm{a}$ and a block diagonal matrix with the submatrices $\bm{A}_1$ and $\bm{A}_2$, respectively. $\bm{I}$ is an identity matrix and $\bm{I}_a$ is an anti-diagonal identity matrix. $\text{T}(\bm{u})$ is a Hermitian Toeplitz matrix with first column being $\bm{u}$. $\|\bm{A}\|_F$, $\bm{A}^\dag$, and $\text{tr}(\bm{A})$ are the Frobenius norm, the pseudoinverse, and the trace of $\bm{A}$, respectively. The operation $\text{vec}(\cdot)$ stacks all the columns of a matrix into a vector. $\otimes$ is the Kronecker product of matrices or vectors. $\mathbb{E}\{\cdot\}$ denotes expectation.

\section{Models and Preliminaries}\label{sect:modelpriorart}
In this section, we first present the signal model and state the goal of both CCE and CSI estimation. Then, we briefly overview the related prior work on relevant super-resolution techniques that are only applicable for ideal uniform arrays, e.g., the uniform linear array (ULA), and under a fixed hybrid hardware structure.

\subsection{Channel and Signal Models}\label{sect:model}
\begin{figure}[!t]
	\centering	\includegraphics[width=3.5in]{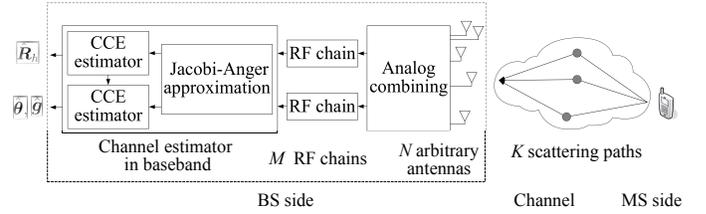}
		\caption{System model of sparse channel estimation in hybrid mmWave massive MIMO systems with arbitrary arrays.}
		\label{fig:system_model}
\end{figure}

Consider a narrowband\footnote{In a wideband case with frequency selectivity, the continuous-valued delays of the individual paths of the sparse time-dispersive channels can be estimated via gridless CS to achieve super-resolution accuracy in the time domain \cite{Pejoski2015Estimation}.} mmWave massive MIMO time division duplex (TDD)\footnote{This work can 
be applied to frequency division duplex (FDD) systems as well, given the angle reciprocity between uplink and downlink \cite{Xie2017Unified}.} system for channel estimation conducted at the base station (BS). As shown in Fig.~\ref{fig:system_model}, 
the BS has a hybrid structure equipped with $N$ arbitrarily deployed antennas and $M \,(M<N)$ RF chains. For simplicity, 
we mainly focus on the basic single-antenna case at the mobile station (MS), while we extend to the case of multiple-antenna 
MS as well in Section~\ref{sect:multi-MS-antenna}.
Noticeably, as shown in Fig.~\ref{fig:system_model}, to impose the useful Vandermonde structure in an arbitrary array geometry, a preprocessing block via Jacobi-Anger approximation is added to the channel estimator in the hybrid mmWave massive MIMO system, which will be described in Section~\ref{sect:JAapprox}.

At the mmWave frequency, the wireless channel experiences limited scattering propagation, which results in a sparse multipath structure \cite{Shafi2018Microwave, Rappaport2014Millimeter}, as shown in Fig.~\ref{fig:system_model}. In this sense, the channel can be described by a geometric model with $K \,(K< M<N)$ scatterers\footnote{In this work, we focus on the case of point scatterers. The angle spread issue due to the reflecting areas of shaped scatterers is out of scope of this paper. The impact of angle spreads on the proposed methods will be studied in future work.}, in which each path is parameterized by the path angle and the path gain. For simplicity, suppose that each scatterer contributes to one propagation path, which can be straightforwardly extended to 
cluster scattering 
where each cluster includes multiple scattering paths \cite{Ayach2014Spatially}. Further, in mmWave channels, the angles of the scattering paths remain constant for a relatively long time, while the channel coefficients change very rapidly \cite{Rappaport2014Millimeter}. Accordingly, the uplink channel $\bm{h}_t$ can be expressed as the sum of $K$ paths in the form
\begin{align}\label{eq:channel}
\begin{split}
\bm{h}_t=\sum_{k=1}^K g_{k,t}\bm{a}(\theta_k), \ \forall t,
\end{split}
\end{align}
where $g_{k,t}$ denotes the channel gain for the $k$-th scattering path at the $t$-th snapshot, and $\bm{a}(\theta_k)\in\mathbb{C}^{N}$ is the array manifold vector corresponding to the $k$-th channel path.

In this work, we focus on arbitrary arrays, in which the $n$-th antenna element is placed at a known location $(r_n,\phi_n), n=1,\dots,N$ in 
polar coordinates. Define $\theta_k$ as the angle between the polar axis and the $k$-th path, and take the polar origin as the reference point. Then, the $n$-th component of the array manifold vector for the $k$-th path can be written as
\begin{align}\label{eq:arbitrary}
\begin{split}
[\bm{a}(\theta_k)]_n=e^{j2\pi\frac{r_n}{\lambda}\cos(\theta_k-\phi_n)},
\end{split}
\end{align}
where $\lambda$ denotes the wavelength. In a compact matrix-vector form, the channel $\bm{h}_t$ in \eqref{eq:channel} can be rewritten as
\begin{equation}\label{eq:h=Ag}
  \bm{h}_t=\bm{A}\bm{g}_t,
\end{equation}
where $\bm{g}_t=[g_{1,t},\dots,g_{K,t}]^T$ and $\bm{A}=[\bm{a}(\theta_1),\dots,\bm{a}(\theta_K)]$.

In 
uplink channel estimation, the MS sends out training symbols $z_t$ which are also known to the BS. For simplicity, let $|z_t|=1$ for all snapshots. Then, the received signal at the BS's antennas can be represented as
\begin{align}\label{eq:RecSigAntennas}
\begin{split}
\bm{x}_t=\bm{h}_t z_t +\bm{w}_t=\bm{A}\bm{g}_t z_t +\bm{w}_t,
\end{split}
\end{align}
where $\bm{w}_t$ denotes additive Gaussian noise distributed as $\mathcal{CN}(\bm{0},\sigma^2\bm{I})$.
According to \eqref{eq:RecSigAntennas}, the covariance matrices for $\bm{x}_t, \bm{h}_t$ and $\bm{g}_t$ have the following linear relationship:
\begin{align}\label{eq:Rx-Rh-Rg}
\begin{split}
\bm{R}_x=\mathbb{E}\{\bm{x}_t\bm{x}_t^H\}=\bm{R}_h+\sigma^2\bm{I}=\bm{A}\bm{R}_g\bm{A}^H+\sigma^2\bm{I},
\end{split}
\end{align}
where $\bm{R}_h=\mathbb{E}\{\bm{h}_t\bm{h}_t^H\}$, and $\bm{R}_g=\mathbb{E}\{\bm{g}_t\bm{g}_t^H\}$.

The BS adopts a hybrid hardware structure in the form of $\bm{W}_t = \bm{W}^{\text{BB}}_t \bm{W}^{\text{RF}}_t$, where $\bm{W}^{\text{BB}}_t \in\mathcal{C}^{M\times M}$ denotes a baseband digital combiner, and $\bm{W}^{\text{RF}}_t \in\mathcal{C}^{M\times N}$ is an analog combiner. In this paper, we focus on the case where $\bm{W}^{\text{RF}}_t$ is made of a network of random phase shifters, while this work can be applied to other structures such as the switch-based network as well \cite{Mendez-Rial2015Channel}. To further enhance randomness, $\bm{W}^{\text{BB}}_t$ can be set as a random Gaussian matrix.
After being multiplied with the hybrid combining matrix $\bm{W}_t\in\mathcal{C}^{M\times N}$ and the known 
training symbol $z_t^*$, the received signal at the lower-dimensional baseband is given by
\begin{align}\label{eq:RecSigBBy}
\begin{split}
\bm{y}_t= z_t^* \bm{W}_t\bm{x}_t=\bm{W}_t\bm{A}\bm{g}_t+\bm{W}_t\bm{n}_t,
\end{split}
\end{align}
where $\bm{n}_t = z_t^* \bm{w}_t$. When $\bm{W}_t$ in \eqref{eq:RecSigBBy} is different snapshot by snapshot, it is a dynamic channel sensing system. 

\begin{figure}[!t]
	\centering	\includegraphics[width=3.5in]{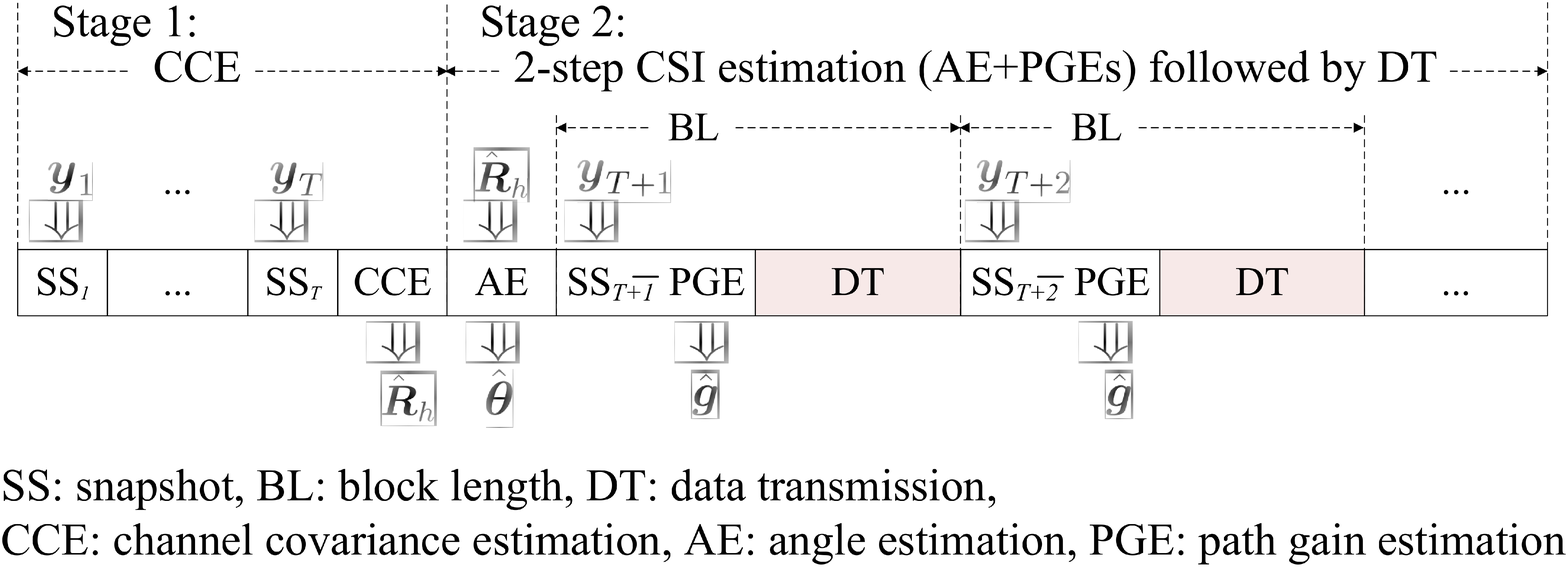}
		\caption{Illustration of two-stage channel estimation including both channel covariance estimation and instantaneous CSI estimation that is further divided into angle estimation and path gain estimation.}
		\label{fig:2stage2step-illustration}
\end{figure}

In this paper, we assume a 
block fading channel, where path gains, and hence the CSI, stay constant within a block but vary 
from block to block. In contrast, path angles vary much slower, and stay unchanged across blocks, until angle re-calibration is needed.
Further, angles can be retrieved from the channel covariance, which is the key idea behind the statistical inference methods for angle
estimation.
This motivates us to design a two-stage channel estimation framework to obtain both the channel covariance and the instantaneous CSI sequentially, as shown in Fig.~\ref{fig:2stage2step-illustration}. In the first stage, we apply CCE to obtain the channel covariance over multiple snapshots. Then, in the second stage, considering the difference in time-variation of path angles and path gains, we design a two-step scheme to do instantaneous CSI estimation, which is further divided into two subproblems: angle estimation and path gain estimation.

\noindent {\bf \em Remark 1:} When the goal of channel estimation is CCE only, the estimator can terminate upon completing the first stage. The CCE by itself is relevant in two cases: one is to simplify either the channel estimation task or the transceiver design, and the other is when path gains experience fast fading that renders the CSI estimates useless for data transmission. In both cases, CCE-based transceiver design can be adopted \cite{Park2017Exploiting, Li2017Optimizing}.

\noindent {\bf \em Remark 2:} Our task of CCE-based angle estimation is also useful during the system calibration stage for fixed wireless applications, in which case the estimated angles can be used to facilitate several system-level tasks such as user grouping and beam-sectoring.

\subsection{Prior Art: Efficient Super-Resolution Techniques in Ideal Models}
When the block length is not long enough for traditional super-resolution methods to work for massive MIMO when $N$ is very large, we have to focus on those super-resolution techniques that exploit the structural feature of the array manifold to improve the sample efficiency.
Therefore, in this subsection, we overview existing efficient super-resolution techniques for channel estimation \cite{Wang2017Efficient, Tsai2018Millimeter}, which critically rely on an assumption that the antenna array has to be a uniform array, e.g., a ULA. Then, the array manifold naturally presents an explicit Vandermonde structure. That is, the $n$-th component of the array manifold vector at the $k$-th path is a special case of \eqref{eq:arbitrary} with $\phi_n=0, \, r_n=d(n-1), \forall n$, which has the form
\begin{align}
\begin{split}
[\bm{a}(\theta_k)]_n=e^{j2\pi(n-1)\frac{d}{\lambda}\cos(\theta_k)},
\end{split}
\end{align}
where $d$ denotes the same separation distance between any two adjacent antennas placed along the polar axis. Without loss of generality, suppose $d=\lambda/2$.

Further, assuming $\bm{W}_t=\bm{W}, \, \forall t$ as the fixed hybrid structure over time, the received signals $\bm{y}_t$ for $t=1,\dots,T$ in \eqref{eq:RecSigBBy} can be collected to form a matrix $\bm{Y}$ with $\bm{y}_t$ being its columns as
\begin{align}\label{eq:Y}
\begin{split}
\bm{Y}=\bm{W}\bm{A}\bm{G}+\bm{W}\bm{N}=\bm{W}\bm{H}+\bm{W}\bm{N},
\end{split}
\end{align}
where the matrices $\bm{H}$, $\bm{G}$ and $\bm{N}$ are similarly defined as $\bm{Y}$ with $\bm{h}_t$, $\bm{g}_t$ and $\bm{n}_t$ being their columns, respectively. From \eqref{eq:Y}, the covariance of $\bm{y}_t$ is given as
\begin{align}
\begin{split}\label{eq:idealRy}
\bm{R}_y=\mathbb{E}\{\bm{y}_t\bm{y}_t^H\}=\bm{W}(\bm{R}_h+\sigma^2\bm{I})\bm{W}^H.
\end{split}
\end{align}

Then, an atom set in the MMV case is defined as \cite{Li2016Offgrid,Yang2016Exact}
\begin{align}
\begin{split}\label{eq:idealatomset}
\mathcal{A}=\left\{\bm{a}(f)\,\bm{b}^H \left| \; f\in\left(\textstyle-\frac{1}{2}, \frac{1}{2}\right], \bm{b}\in\mathbb{C}^T, \normt{\bm{b}}=1 \right.\right\},
\end{split}
\end{align}
where $\bm{a}(f)\in\mathbb{C}^N$ with the $n$-th component being $e^{j2\pi(n-1) f}$.
According to the atomic norm theorem \cite{Li2016Offgrid,Yang2016Exact}, the atomic norm of $\bm{H}$ over the atom set $\mathcal{A}$ is defined as
\begin{align}\label{eq:atomicnorm}
\begin{split}
\norma{\bm{H}}=\inf\left\{l>0:\bm{H}\in l \, \text{conv}(\mathcal{A})\right\},
\end{split}
\end{align}
which seeks the most concise representation of $\bm{H}$ by involving the fewest atoms over $\mathcal{A}$.

From the received signals $\bm{Y}$ in \eqref{eq:Y}, the channel estimation for the instantaneous CSI is conducted by solving the regularized ANM formulation as
\begin{align}\label{eq:idealANM}
\begin{split}
\hat{\bm{H}}=\arg\min_{\bm{H}} \norma{\bm{H}}+\frac{\tau}{2}\normf{\bm{Y}-\bm{W}\bm{H}},
\end{split}
\end{align}
where $\tau$ denotes the regularization coefficient controlling the tradeoff between the ANM and the residual error tolerance to the observations. According to \cite{Bhaskar2013Atomic},  $\tau$ can be set as $\tau = {1}/{(\sigma +\frac{\sigma}{\log N}{\sqrt{N\log N+N\log(4\pi \log N)}}\, )}$.

Besides the instantaneous CSI itself, the second-order channel statistics in terms of the channel covariance $\bm{R}_h$ also play an important role in precoding design for mmWave massive MIMO channels \cite{Park2017Exploiting, Li2017Optimizing}. Suppose the channel gains of the sparse paths are uncorrelated with each other. Then, $\bm{R}_h$ not only presents the low rankness due to the channel sparsity, but also is a well-structured positive semidefinite (PSD) Hermitian Toeplitz matrix thanks to the Vandermonde structure of uniform arrays.

To utilize these useful features in the channel statistics, a structure-based optimization approach named low-rank structured matrix 
reconstruction (LRSMR) \cite{Li2016Offgrid,Wu2017Toeplitz}, can be applied to do CCE, by imposing the Hermitian Toeplitz structure on $\bm{R}_h$ as $\bm{R}_h=\text{T}(\bm{u}_h)$. Thus, from the sample covariance matrix $\hat{\bm{R}}_y=\frac{1}{T}\bm{Y}\bm{Y}^H$, the CCE can be conducted via LRSCR as,
\begin{align}\label{eq:IdealLRSCR}
\begin{split}
\hat{\bm{R}}_h=&\arg\min_{\text{T}(\bm{u}_h)} \text{tr}\left(\text{T}(\bm{u}_h)\right)\!+\!\frac{\tau}{2}\normf{\hat{\bm{R}}_y\!-\!\bm{W}\text{T}(\bm{u}_h)\bm{W}^H}\\
&~~~~\text{s.t.}~~~~\text{T}(\bm{u}_h)\succeq0.
\end{split}
\end{align}

Unfortunately, the uniform array assumption cannot be guaranteed in practice, considering the antenna misalignment and subarray selection issues arisen in hybrid mmWave massive MIMO systems. Moreover, the assumption of the fixed combining matrix over all snapshots is ineffective to find out all potential scattering paths in mmWave channel estimation. Regarding these practical situations, two questions arise: 1) can we design new super-resolution channel estimation techniques via LRSCR and ANM for arbitrary arrays? 2) how can we retrieve the desired channel information via the dynamic hybrid structure? In the next section, to fully address these problems, we develop super-resolution and fast channel estimation approaches for obtaining both the channel covariance and the instantaneous CSI.

\section{Super-Resolution Channel Estimation for Arbitrary Arrays}\label{sect:superChEst}
In this section, we first leverage the Jacobi-Anger approximation to extract the useful Vandermonde structure from a nonideal array geometry. Then, two new CCE methods are developed through structure-based optimization techniques for arbitrary arrays. For efficient estimation of the instantaneous CSI, a two-step solution is proposed, which estimates the path angles and path gains sequentially.

\subsection{Jacobi-Anger Approximation}\label{sect:JAapprox}
The Jacobi-Anger expansion provides a general infinite terms expansion of exponentials of trigonometric functions in the basis of their harmonics \cite{Abramowitz1964Handbook}. Specifically, \eqref{eq:arbitrary} in the Jacobi-Anger expansion form is expressed as
\begin{align}\label{eq:JA-expansion}
\begin{split}
[\bm{a}(\theta_k)]_n=\sum_{i=-\infty}^{+\infty}j^i\text{J}_i\left(2\pi \frac{r_n}{\lambda}\right)e^{-j\phi_n i}e^{j\theta_k i},
\end{split}
\end{align}
where $\text{J}_i(\cdot)$ denotes the $i$-th order Bessel function of the first kind.

Although \eqref{eq:JA-expansion} indicates a summation of infinite series, the value of $|\text{J}_i(r)|$ decays very rapidly as the value of $|i|$ increases for any $r>0$, which is a nature of the Bessel function. Thus, the infinite series expansion can be well approximated by keeping only the terms having large absolute values located around the central range of the series, that is, $|i|\leq I$ in \eqref{eq:JA-expansion}. To reach a desired precision, the maximum mode order $I$ for the approximation is chosen as \cite{Mathews1994Eigenstructure}
\begin{align}\label{eq:modeNum1}
\begin{split}
I > \frac{2\pi}{\lambda}r_{max},
\end{split}
\end{align}
where $r_{max}$ is the maximum $r_n$.

Then, given $I$, \eqref{eq:JA-expansion} can be approximately expressed as
\begin{align}\label{eq:JA-approx}
\begin{split}
[\bm{a}(\theta_k)]_n&\approx\sum_{i=-I}^{I}j^i\text{J}_i(2\pi \frac{r_n}{\lambda})e^{-j\phi_n i}e^{j\theta_k i}\\
&=\bm{c}_n^T\bm{v}(\theta_k),
\end{split}
\end{align}
where $\bm{c}_n$ and $\bm{v}(\theta_k)$ are given by
\begin{subequations}
    \begin{align}
      &[\bm{c}_n]_i= j^i\text{J}_i(2\pi \frac{r_n}{\lambda})e^{-j\phi_n i},  \; i=-I,\ldots,I;\label{eq:c-i} \\
      &[\bm{v}(\theta_k)]_i= e^{j\theta_k i}, \; i=-I,\ldots,I.\label{eq:v-i}
    \end{align}
\end{subequations}
According to \eqref{eq:JA-approx}, the $N \times 1$ array manifold vector of an arbitrary array can be approximated as
\begin{align}\label{eq:a-C-v}
\begin{split}
\bm{a}(\theta_k)=[\bm{c}_1,\dots,\bm{c}_N]^T\bm{v}(\theta_k)=\bm{C}\bm{v}(\theta_k).
\end{split}
\end{align}

Noticeably, it is the Jacobi-Anger expansion that enables to separate the unknown channel-related parameter ($\theta_k$) in \eqref{eq:v-i} from the known array-related configurations ($r_n$ and $\phi_n$) in \eqref{eq:c-i}, which are otherwise mingled in the original physical antenna domain of \eqref{eq:arbitrary}. Further, due to the exponential form in \eqref{eq:v-i}, the Vandermonde structure is well-presented in the virtual uniform antenna domain, in terms of the virtual array manifold
\begin{equation}\label{eq:V}
  \bm{V}\left(\bm{\theta}\right)=\left[\bm{v}(\theta_1),\dots,\bm{v}(\theta_K)\right].
\end{equation}
Merging \eqref{eq:a-C-v} and \eqref{eq:V}, any arbitrary array can be expressed as the multiplication of a Bessel matrix and a Vandermonde matrix, in the form
\begin{equation}\label{eq:A=CV}
  \bm{A}=\bm{CV}.
\end{equation}

Therefore, to appreciate the Vandermonde structure, we reformulate the channel by substituting \eqref{eq:a-C-v} into \eqref{eq:h=Ag} as
\begin{align}\label{eq:h=Cd}
\begin{split}
\bm{h}_t&=\bm{C}\bm{V}\bm{g}_t=\bm{C}\bm{d}_t,
\end{split}
\end{align}
where $\bm{d}_t=\bm{V}\bm{g}_t$ represents the virtual channel in the virtual uniform antenna domain. From \eqref{eq:Rx-Rh-Rg} and \eqref{eq:A=CV}, the channel covariance can be rewritten as
\begin{align}\label{eq:Rh}
\begin{split}
\bm{R}_h&=\bm{C}\bm{V}\bm{R}_g\bm{V}^H\bm{C}^H=\bm{C}\bm{R}_v\bm{C}^H,
\end{split}
\end{align}
where $\bm{R}_v$ denotes the virtual channel covariance matrix given by
\begin{equation}\label{eq:Rv}
  \bm{R}_v= \bm{V}\bm{R}_g\bm{V}^H.
\end{equation}

Accordingly, after taking the hybrid structure and arbitrary array into consideration, the received signal in \eqref{eq:RecSigBBy} can be expressed as
\begin{align}
\begin{split}
\bm{y}_t&=\bm{W}_t\bm{C}\bm{V}\bm{g}_t+\bm{W}_t\bm{n}_t\\
&=\bm{\Phi}_t\bm{d}_t+\bm{W}_t\bm{n}_t,
\end{split}
\end{align}
where $\bm{\Phi}_t=\bm{W}_t\bm{C}$ denotes the equivalent sensing matrix.

As the Bessel matrix $\bm{C}$ depends solely on the known array geometry, $\bm{R}_h$ in \eqref{eq:Rh} can be estimated as long as $\bm{R}_v$ is retrieved. Next, we need to figure out how to estimate $\bm{R}_v$ from collected $\{\bm{y}_t\}_t$.

\subsection{Channel Covariance Estimation}\label{sect:CCE}
In this subsection, to estimate the virtual channel covariance $\bm{R}_v$, and hence also the actual channel covariance $\bm{R}_h$, we develop two super-resolution CCE techniques for the MMV case with multiple snapshots. One is the CCE via LRSCR, and the other is the CCE via ANM. The ANM-based CCE method is applicable to the dynamic hybrid hardware structure.

\subsubsection{CCE via LRSCR}\label{sect:LRSCR}
Suppose the path gains of the fading channels are uncorrelated with each other. Then, $\bm{R}_v$ in \eqref{eq:Rv} not only presents the low rankness because of sparse scattering environments, but also is a well-structured PSD Hermitian Toeplitz matrix due to the Vandermonde structure of the virtual uniform array. In this sense, the LRSCR technique can be used to implement the low-rank feature of $\bm{R}_v$.
Moreover, suppose the hybrid combining matrix $\bm{W}_t$ is fixed over time, i.e., $\bm{W}_t = \bm{W}$, and $\bm{\Phi}_t=\bm{WC}=\bm{\Phi},  \forall t$. Substituting \eqref{eq:Rh} into \eqref{eq:idealRy}, $\bm{R}_y$ can then be rewritten as
\begin{align}
\begin{split}
{\bm{R}}_y&=\bm{W}\bm{C}\bm{R}_v\bm{C}^H\bm{W}^H+\sigma^2\bm{W}\bm{W}^H\\
&=\bm{\Phi}\bm{R}_v\bm{\Phi}^H+\sigma^2\bm{W}\bm{W}^H,
\end{split}
\end{align}
which clearly is a function of $\bm{R}_v$.

Accordingly, by imposing the Hermitian Toeplitz structure on $\bm{R}_v$ in terms of $\bm{R}_v=\text{T}(\bm{u}_v)$, we can describe the LRSCR-based formulation for the estimation of $\bm{R}_v$ from the sample covariance $\hat{\bm{R}}_y$ as
\begin{align}\label{eq:LRSCR}
\begin{split}
\hat{\bm{R}}_v=&\arg\min_{\text{T}(\bm{u}_v)} \text{tr}\left(\text{T}(\bm{u}_v)\right)\!+\!\frac{\tau}{2}\normf{\hat{\bm{R}}_y\!-\!\bm{\Phi}\text{T}(\bm{u}_v)\bm{\Phi}^H}\\
&~~~~\text{s.t.}~~~~\text{T}(\bm{u}_v)\succeq0.
\end{split}
\end{align}
Then, given $\hat{\bm{R}}_v$ estimated from \eqref{eq:LRSCR}, we finally obtain $\hat{\bm{R}}_h$ via \eqref{eq:Rh}, with known $\bm{C}$.

Further, to solve \eqref{eq:LRSCR} with lower computational complexity compared with using off-the-shelf SDP solvers \cite{CVX}, we will develop a fast algorithm via ADMM later in Section~\ref{sect:ADMM}.

It is worth noting that CCE via LRSCR can be done in blind mode from data symbols, since it can work as long as $\hat{\bm{R}}_y$ is available to \eqref{eq:LRSCR}.

\subsubsection{CCE via Dynamic-ANM}\label{sect:DynamicANM}
Let $\bm{D}=[\bm{d}_1,\dots,\bm{d}_T]$ collect the virtual channels defined in \eqref{eq:h=Cd} from different time slots. Then, an atom set can be defined in the virtual antenna domain as
\begin{align}\label{eq:atomset}
\begin{split}
\mathcal{A}^\prime=\left\{\bm{v}(f)\bm{q}^H \left| \, f\in(\textstyle-\frac{1}{2},\frac{1}{2}],\bm{q}\in\mathbb{C}^T,\normt{\bm{q}}=1 \right.\right\},
\end{split}
\end{align}
where $\bm{v}(f)\in\mathbb{C}^{2I+1}$ with its components being $e^{j2\pi i f}, \, i=-I,\ldots, I$. Obviously, $\bm{D}$ is a linear combination of the atoms from the set $\mathcal{A}^\prime$. In this sense, the ANM technique can be used to exploit the low rank and Vandermonde features of $\bm{D}$.

Then, with the received signals $\{\bm{y}_t\}_t$ in the matrix-form $\bm{Y}$, we produce the following ANM formulation:
\begin{align}\label{eq:FixANM}
\begin{split}
\hat{\bm{D}}=\arg\min_{\bm{D}} \|\bm{D}\|_{\mathcal{A}^\prime}+\frac{\tau}{2}\normf{\bm{Y}-\bm{\Phi}\bm{D}}.
\end{split}
\end{align}

Next, we consider a 
time-varying $\bm{\Phi}_t$ in the dynamic hybrid structure case. Since the ANM in our objective function can be maintained, we now only have to formulate the residual error tolerance snapshot by snapshot. Hence, the dynamic version of \eqref{eq:FixANM} can be rewritten as
\begin{align}\label{eq:DynamicANM}
\begin{split}
\hat{\bm{D}}=\arg\min_{\bm{D}} \|\bm{D}\|_{\mathcal{A}^\prime}+\frac{\tau}{2}\sum_{t=1}^T\normf{\bm{y}_t-\bm{\Phi}_t\bm{d}_t}.
\end{split}
\end{align}

Further, replacing the atomic norm in \eqref{eq:DynamicANM} by its SDP formulation \cite{Li2016Offgrid}\cite{Yang2016Exact}, we have
\begin{align}\label{eq:DynamicANMSDP}
\begin{split}
(\hat{\bm{D}},&\text{T}(\hat{\bm{u}}_d),\hat{\bm{Q}})=\\
\arg&\min_{\bm{D},\text{T}(\bm{u}_d),\bm{Q}} ~~~\frac{1}{2\sqrt{2I+1}}\left(\text{tr}\left(\text{T}\left(\bm{u}_d\right)\right)+\text{tr}\left(\bm{Q}\right)\right) \\
&+\frac{\tau}{2}\sum_{t=1}^T\normf{\bm{y}_t-\bm{\Phi}_t\bm{d}_t},
~~~\text{s.t.}~\left[\begin{array}{cc}
\text{T}(\bm{u}_d)&\bm{D}\\
\bm{D}^H&\bm{Q}
\end{array}\right]\succeq 0.\\
\end{split}
\end{align}
For solving the problem \eqref{eq:DynamicANMSDP}, an ADMM-based fast algorithm can also be designed and iteratively implemented in a similar way as for \eqref{eq:LRSCR}.

In addition, according to the atomic norm definition and its SDP formulation, the reconstructed $\text{T}(\hat{\bm{u}}_d)$ from \eqref{eq:DynamicANMSDP} can be expressed by a Vandermonde decomposition as
\begin{align}\label{eq:T=VRg2V}
\begin{split}
\text{T}(\hat{\bm{u}}_d)=\hat{\bm{V}}\left(T\hat{\bm{R}}_g\right)^{\frac{1}{2}}\hat{\bm{V}}^H.
\end{split}
\end{align}

Finally, using the estimated $\hat{\bm{V}}$ and $\hat{\bm{R}}_g$, we can obtain $\hat{\bm{R}}_v$ via \eqref{eq:Rv}  and then $\hat{\bm{R}}_h$ via \eqref{eq:Rh} accordingly.

\subsection{Two-Step Instantaneous CSI Estimation}\label{sect:CSI}
As has been mentioned in Section~\ref{sect:model}, the spatially sparse mmWave channel $\bm{h}$ is fully determined by parameters: $\bm{\theta}$ and $\bm{g}$.
Since the path angles depend only on the relative positions of the BS, the MS, and the scatterers,
$\bm{\theta}$ varies 
slowly, but can take a long time to acquire accurately from $\hat{\bm{R}}_h$ according to statistical inference techniques for angle estimation \cite{Schmidt1986Multiple, Roy1989ESPRIT, Hua1990Matrix}. In contrast, the path gains $\bm{g}$ are easy to acquire given $\hat{\bm{\theta}}$, but vary frequently.
Therefore, given the estimated $\hat{\bm{R}}_h$, in the second stage, we design a two-step CSI estimation scheme for block transmission.
Angle estimates directly result from the CCE output, which stay unchanged for multiple blocks until they change. Given $\hat{\bm{\theta}}$, path gains are estimated at each block, followed by CSI-based data transmission.

\subsubsection{Angle Estimation}\label{sect:AngleEst}
Thanks to the specific matrix structures presented by the second-order statistics of mmWave channels, such as the low rankness due to the sparse scattering propagation and the PSD Toeplitz structure imposed by the SDP formulation through either \eqref{eq:LRSCR} or \eqref{eq:DynamicANMSDP}, the Vandermonde decomposition can be applied to estimate the angles.
\begin{itemize}
  \item If the LRSCR technique is used for CCE in Section~\ref{sect:LRSCR}, the recovered $\hat{\bm{R}}_v$ from \eqref{eq:LRSCR} is a low-rank PSD Toeplitz matrix. According to the Vandermonde decomposition lemma \cite{Toeplitz1911Zur}, $\hat{\bm{R}}_v$ can be uniquely expressed as
      \begin{equation}\label{eq:Rv=VRgV}
        \hat{\bm{R}}_v = \hat{\bm{V}} \hat{\bm{R}}_g \hat{\bm{V}}^H.
      \end{equation}
      This Vandermonde decomposition can be computed efficiently via root finding or by solving a generalized eigenvalue problem \cite{Hua1990Matrix}.
      Since the virtual uniform array geometry is solely decided by the angles as in \eqref{eq:V}, $\hat{\bm{\theta}}$ can be directly extracted from $\hat{\bm{V}}$ according to \eqref{eq:v-i}.
  \item For the CCE based on the Dynamic-ANM technique as developed in Section~\ref{sect:DynamicANM}, the Vandermonde decomposition can be carried out as in \eqref{eq:T=VRg2V}. Although the definition of the atom set in \eqref{eq:atomset} leads to a diagonal matrix in the form of $(T\hat{\bm{R}}_g)^{\frac{1}{2}}$ in \eqref{eq:T=VRg2V} that is different from $\hat{\bm{R}}_g$ in \eqref{eq:Rv=VRgV}, the common Vandermonde structure of $\hat{\bm{V}}$ still leads to the same estimation results for $\hat{\bm{\theta}}$ via \eqref{eq:v-i}.
\end{itemize}

\subsubsection{Path Gain Estimation}
Given the obtained angular information $\hat{\bm{\theta}}$ from Section~\ref{sect:AngleEst}, next we need to estimate the path gains $\bm{g}$ in a timely fashion as shown in Fig.~\ref{fig:2stage2step-illustration}. To this end, we first form the array matrix $\bm{A}$ via \eqref{eq:arbitrary}. Then, following the principle of a matched filter, we tune the precoder as $\bm{W}=\bm{A}^H$ for beamforming.
Noteworthily, with the obtained $\hat{\bm{\theta}}$ from angle estimation, the $K \times N$ matched filter based $\bm{W}$ applied now for path gain estimation is different from the $M \times N$ random phase shifter based $\bm{W}$ used earlier for CCE when $\hat{\bm{\theta}}$ is unknown.
As a result, the estimation of $\bm{g}$ can be expressed as a least squares formulation
\begin{equation}\label{eq:LS}
  \hat{\bm{g}} =  \arg\min_{\bm{g}} \normt{\bm{y} - \bm{W} \bm{A} \bm{g}} = \left(\bm{WA}\right)^\dag \bm{y}.
\end{equation}

\section{Discussion of Related Issues}
\label{sect:discussions}
In this section, we provide detailed discussions on some specific issues related to the proposed solutions.
We first provide the theoretical results in terms of fundamental limits for the proposed super-resolution channel estimation in hybrid mmWave massive MIMO with arbitrary arrays. Then, we design a first-order algorithm via ADMM to rapidly 
implement the super-resolution estimation in lieu of invoking the high-computational SDP. In addition, we study the spurious-peak issue of the Jacobi-Anger approximation as a side effect specific to 
nonuniform linear arrays, and provide an effective way to solve this problem. Finally, we extend the work to the multiple-antenna MS 
case, which is developed based on an efficient 2D gridless CS approach.

\subsection{Analysis of Fundamental Limits}
To get an approximation with certain precision, \eqref{eq:modeNum1} offers a lower bound on the choice of the maximum mode order $I$ of the Jacobi-Anger expansion. From the view of the Jacobi-Anger approximation, the larger $I$ is, the higher the accuracy the approximation can achieve. On the other hand, from the view of sparse channel estimation, we need to retrieve the sparse virtual channel $\{\bm{d}_t\}_t$ from the compressed measurements $\{\bm{y}_t\}_t$ with a high probability. For a given hybrid mmWave massive MIMO system, the dimension of $\bm{y}_t$ is fixed and known as $M$. To ensure the proposed gridless CS based methods are feasible and effective, first, it is necessary to study the minimum number of RF chains $M_{min}$ required by our approaches given the approximation and hybrid structure.
According to Theorem III.4 in \cite{Li2018Atomic}, to guarantee successful reconstruction of channel covariance with high probability, $M_{min}$ can be expressed as a function of $I$ and $K$ as
\begin{align}\label{eq:Mmin}
\begin{split}
M_{min} = C K \log(2I+1),
\end{split}
\end{align}
where $C$ is a numerical constant. Note that although the sensing matrix $\bm{\Phi}_t$ in \cite{Li2018Atomic} is assumed to be an i.i.d. random Gaussian matrix, it is reasonable to relax \eqref{eq:Mmin} to accommodate the analysis in this paper when $\bm{\Phi}_t$ is modeled as a random phase shifter based $\bm{W}_t$ multiplied by a Bessel matrix $\bm{C}$. Then, replacing $M_{min}$ in \eqref{eq:Mmin} by $M$ with $M > M_{min}$, we can obtain an upper bound on 
$I$.
Combining with \eqref{eq:modeNum1}, we have
\begin{align}\label{eq:lower-I-upper}
\begin{split}
\frac{2\pi}{\lambda}r_{max} < I < \frac{1}{2}\left(e^{\frac{M}{CK}}-1\right).
\end{split}
\end{align}
It is worth noting that \eqref{eq:lower-I-upper} actually reflects the tradeoff between the approximation accuracy to combat the imperfection of the array geometry and the hardware cost required for sparse channel estimation, which thus sheds light on the choices of $I$ and $M$ in practice.

\subsection{Fast Implementation via ADMM}\label{sect:ADMM}
To avoid the high computational complexity of the SDP-based solutions, we develop a fast iterative algorithm via ADMM. Next, we mainly discuss the solution for the LRSCR formulation \eqref{eq:LRSCR} and omit that for the ANM case \eqref{eq:DynamicANMSDP}\footnote{The design of an ADMM-based fast algorithm for \eqref{eq:DynamicANMSDP} can be developed in a similar way as for \eqref{eq:LRSCR}. The 
differences are the two additional variables $\bm{D}$ and $\bm{Q}$ in \eqref{eq:DynamicANMSDP}, which can be easily updated by gradient descent in each iteration. Moreover, the algorithm implementation for \eqref{eq:DynamicANMSDP} is actually simpler than that for \eqref{eq:LRSCR}. The reason is that the Toeplitz structured matrix $\text{T}(\bm{u})$ is not included in the least squares term in the objective function in \eqref{eq:DynamicANMSDP}, which then simplifies the calculation of the partial derivative of the augmented Lagrangian $\mathcal{L}$ with respect to $\bm{u}^*$. As a result, compared with the LRSCR case, the update of $\bm{u}$ in each iteration becomes easier in the ANM case.}. To apply ADMM \cite{Boyd2010Distributed}, we reformulate \eqref{eq:LRSCR} as 
\begin{align}
\begin{split}
\hat{\bm{R}}_v=&\arg\min_{\text{T}(\bm{u}_v)} \text{tr}\left(\text{T}(\bm{u}_v)\right)\!+\!\frac{\tau}{2}\normf{\hat{\bm{R}}_y\!-\!\bm{\Phi}\text{T}(\bm{u}_v)\bm{\Phi}^H}\\
&~~~~\text{s.t.}~~~~\bm{U}=\text{T}(\bm{u}_v),~\bm{U}\succeq0,
\end{split}
\end{align}
whose augmented Lagrangian can be expressed as
\begin{align}\label{eq:augLag}
\begin{split}
&\mathcal{L}(\bm{u}_v,\bm{U},\bm{\Lambda})\\
=&\text{tr}\left(\text{T}(\bm{u}_v)\right)+\frac{\tau}{2}\normf{\hat{\bm{R}}_y\!-\!\bm{\Phi}\text{T}(\bm{u}_v)\bm{\Phi}^H}
\\&+\langle\bm{\Lambda},\bm{U}-\text{T}(\bm{u}_v)\rangle+\frac{\rho}{2}\normf{\bm{U}-\text{T}(\bm{u}_v)}\\
=&\text{tr}\left(\text{T}(\bm{u}_v)\right)+\frac{\tau}{2}\normf{\hat{\bm{R}}_y\!-\!\bm{\Phi}\text{T}(\bm{u}_v)\bm{\Phi}^H}
-\frac{1}{2\rho}\normf{\bm{\Lambda}}\\&+\frac{\rho}{2}\normf{\bm{U}-\text{T}(\bm{u}_v)+\rho^{-1}\bm{\Lambda}},
\end{split}
\end{align}
where $\bm{U}$ and $\bm{\Lambda}$ are Hermitian matrices. Then the implementation of ADMM involves the following iterative updates:
\begin{align}
\bm{u}_v^{l+1}&=\arg\min_{\bm{u}_v}\mathcal{L}(\bm{u}_v,\bm{U}^l,\bm{\Lambda}^l);\label{eq:teopupdate}\\
\bm{U}^{l+1}&=\arg\min_{\bm{U}\succeq0}\mathcal{L}(\bm{u}_v^{l+1},\bm{U},\bm{\Lambda}^l);\label{eq:Uupdate}\\
\bm{\Lambda}^{l+1}&=\bm{\Lambda}^l+\rho(\bm{U}^{l+1}-\text{T}(\bm{u}_v^{l+1})),
\end{align}
where the superscript $l$ denotes the $l$-th iteration update. In order to implement \eqref{eq:teopupdate}, we take the partial derivative of \eqref{eq:augLag} with respect to $\bm{u}_v^*$ at the $(l\!+\!1)$-th iteration and force it equal to zero. After taking a series of derivations on $\frac{\partial}{\partial\bm{u}_v^*}\mathcal{L}(\bm{u}_v,\bm{U}^l,\bm{\Lambda}^l)\left|_{\bm{u}_v=\bm{u}_v^{l+1}}\right.=0$, we obtain 
\begin{align}
\begin{split}\label{eq:partialdiff}
&\tau\mathcal{G}(\bm{\Phi}^H\bm{\Phi}\text{T}(\bm{u}_v^{l+1})\bm{\Phi}^H\bm{\Phi})+\rho\mathcal{G}(\text{T}(\bm{u}_v^{l+1}))\\&=
\tau\mathcal{G}(\bm{\Phi}^H\hat{\bm{R}}_y\bm{\Phi})+\rho\mathcal{G}(\bm{U}^l+\rho^{-1}\bm{\Lambda}^l)-N_I\bm{e}_1,
\end{split}
\end{align}
where $N_I=2I+1$ is the column (row) size of $\text{T}(\bm{u}_v^{l+1})$, $\bm{e}_1$ is the $N_I$-length vector with only the first element being one, and $\bm{b}=\mathcal{G}(\bm{B})\in\mathbb{C}^{N_I}$ is a mapping from a matrix to a vector where the $n_I$-th element of $\bm{b}$ is the sum of all the elements $\bm{B}_{i,j}$ in $\bm{B}$ satisfying $i-j+1=n_I$. Moreover, denote $\bm{M}$ as the matrix which satisfies $\bm{b}=\mathcal{G}(\bm{B})=\bm{M}\text{vec}(\bm{B})$ and $\bm{\beta}^l=\tau\mathcal{G}(\bm{\Phi}^H\hat{\bm{R}}_y\bm{\Phi})+\rho\mathcal{G}(\bm{U}^l+\rho^{-1}\bm{\Lambda}^l)$, respectively.

Accordingly, we rewrite \eqref{eq:partialdiff} as
\begin{align}
\begin{split}\label{eq:rewrite-partialdiff}
&\tau\bm{M}\text{vec}(\bm{\Phi}^H\bm{\Phi}\text{T}(\bm{u}_v^{l+1})\bm{\Phi}^H\bm{\Phi})+\rho\bm{M}\text{vec}(\text{T}(\bm{u}_v^{l+1}))\\
&=\bm{\beta}^l-N_I\bm{e}_1\\
\Leftrightarrow&\left(\tau\bm{M}\left[(\bm{\Phi}^H\bm{\Phi})^T\otimes(\bm{\Phi}^H\bm{\Phi})\right]+\rho\bm{M}\right)\text{vec}(\text{T}(\bm{u}_v^{l+1}))\\
&=\bm{\beta}^l-N_I\bm{e}_1\\
\Leftrightarrow&\bm{\Pi}\left[\begin{array}{c}
\bm{u}_{\text{R}}^{l+1}\\
\bm{u}_{\text{I}}^{l+1}
\end{array}\right]=\bm{\beta}^l-N_I\bm{e}_1,
\end{split}
\end{align}
where $\bm{u}_v^{l+1}=\bm{u}_{\text{R}}^{l+1}+j[0,({\bm{u}_{\text{I}}^{l+1}})^T]^T$. Since $\text{T}(\bm{u}_v^{l+1})$ is only determined by the real and imaginary parts of $\bm{u}_v^{l+1}$ as $\bm{u}_\text{R}^{l+1}$ and $\bm{u}_\text{I}^{l+1}$, respectively, there exists a fixed matrix $\bm{\Gamma}$ satisfying $\text{vec}(\text{T}(\bm{u}_v^{l+1}))=\bm{\Gamma}[(\bm{u}_{\text{R}}^{l+1})^T,(\bm{u}_{\text{I}}^{l+1})^T]^T$. Moreover, $\bm{\Pi}=\left(\tau\bm{M}\left[(\bm{\Phi}^H\bm{\Phi})^T\otimes(\bm{\Phi}^H\bm{\Phi})\right]+\rho\bm{M}\right)\bm{\Gamma}$. Since $[(\bm{u}_{\text{R}}^{l+1})^T,(\bm{u}_{\text{I}}^{l+1})^T]^T\in\mathbb{R}^{2N_I-1}$, we can rewrite the $N_I$ complex equations of 
\eqref{eq:rewrite-partialdiff} into $2N_I$ real equations as
\begin{align}
\begin{split}
\left[\begin{array}{c}
\text{Real}\{\bm{\Pi}\}\\
\text{Imag}\{\bm{\Pi}\}
\end{array}\right]
\left[\begin{array}{c}
\bm{u}_{\text{R}}^{l+1}\\
\bm{u}_{\text{I}}^{l+1}
\end{array}\right]=
\left[\begin{array}{c}
\text{Real}\{\bm{\beta}^l\}-N_I\bm{e}_1\\
\text{Imag}\{\bm{\beta}^l\}
\end{array}\right].
\end{split}
\end{align}
Hence, the update rule for $\bm{u}_v$ is given by
\begin{align}
\begin{split}
\bm{u}_v^{l+1}&=\bm{u}_{\text{R}}^{l+1}+j
\left[\begin{array}{c}
0\\
\bm{u}_\text{I}^{l+1}
\end{array}\right]\\
\left[\begin{array}{c}
\bm{u}_{\text{R}}^{l+1}\\
\bm{u}_{\text{I}}^{l+1}
\end{array}\right]&=
\left[\begin{array}{c}
\text{Real}\{\bm{\Pi}\}\\
\text{Imag}\{\bm{\Pi}\}
\end{array}\right]^{\dag}
\left[\begin{array}{c}
\text{Real}\{\bm{\beta}^l\}-N_I\bm{e}_1\\
\text{Imag}\{\bm{\beta}^l\}
\end{array}\right].
\end{split}
\end{align}

Let $\Xi^l=\text{T}(\bm{u}_v^{l+1})\!-\!\rho^{-1}\bm{\Lambda}^l
=\bm{E}^l\bm{\Sigma}^l{\bm{E}^l}^H$ be its eigenvalue decomposition, then based on \eqref{eq:augLag} and \eqref{eq:Uupdate}, we have the update of $\bm{U}$ at the $(l\!+\!1)$-th iteration as
\begin{align}
\bm{U}^{l+1}=\bm{E}^l\bm{\Sigma}_+^l{\bm{E}^l}^H,
\end{align}
where $\bm{\Sigma}_+^l$ is obtained by letting all negative eigenvalues of $\bm{\Sigma}^l$ be zero.

The iterative algorithm will stop until both primal and dual residuals satisfy the pre-set tolerance level \cite{Boyd2010Distributed}.

\subsection{Specific Instance of Nonuniform Linear Arrays}
Nonuniform linear arrays yield one case of arbitrary arrays. However, unlike general random distributed antennas, the antenna elements of nonuniform linear arrays are distributed along a line, which results in the Bessel matrix used by the Jacobi-Anger approximation being axial symmetric. 
This axial symmetric characteristic leads to a special issue in 
the implementation of the proposed Dynamic-ANM and LRSCR methods. 
Next, we discuss this specific instance in detail. 

Suppose the locations of antenna elements of a nonuniform linear array are formed as $(r_n,0)$, $n=1,\dots,N$ in polar coordinates. 
Then, \eqref{eq:JA-approx} is rewritten as
\begin{align}
\begin{split}
[\bm{a}(\theta_k)]_n&\approx\sum_{i=-I}^{I}j^i\text{J}_i(2\pi \frac{r_n}{\lambda})e^{j\theta_k i}=\bm{c}_n^T\bm{v}(\theta).
\end{split}
\end{align}
Based on the property of the Bessel function of the first kind that says
\begin{align}
\begin{split}
\text{J}_{-i}(x)=(-1)^n\text{J}_i(x),~\text{for}~ \forall x>0,
\end{split}
\end{align}
we have
\begin{align}\label{eq:c-i=ci}
\begin{split}
[\bm{c}_n]_{-i}&=j^{-i}\text{J}_{-i}\left(2\pi\frac{r_n}{\lambda}\right)\\
&=j^{i}\text{J}_{i}\left(2\pi\frac{r_n}{\lambda}\right)=[\bm{c}_n]_i.
\end{split}
\end{align}
From \eqref{eq:c-i=ci}, the Bessel matrix $\bm{C}$ explicitly holds 
\begin{align}
\begin{split}
\bm{C}\bm{I}_a=\bm{C}.
\end{split}
\end{align}
Accordingly, \eqref{eq:a-C-v} can be expressed as
\begin{align}
\begin{split}\label{eq:A=CrealV}
\bm{a}(\theta)&=\frac{1}{2}\left(\bm{C}+\bm{C}\right)\bm{v}(\theta)
=\frac{1}{2}\left(\bm{C}+\bm{C}\bm{I}_a\right)\bm{v}(\theta)\\
&=\frac{1}{2}\bm{C}\left(\bm{v}(\theta)+\bm{I}_a\bm{v}(\theta)\right)
=\frac{1}{2}\bm{C}\left(\bm{v}(\theta)+\bm{v}^*(\theta)\right)\\
&=\bm{C}\,\text{Real}\left(\bm{v}(\theta)\right).
\end{split}
\end{align}
Then, given \eqref{eq:A=CrealV}, \eqref{eq:h=Cd} can be rewritten as
\begin{align}\label{eq:hreal}
\begin{split}
\bm{h}_t&=\frac{1}{2}\bm{C}(\bm{V}+\bm{V}^*)\,\bm{g}_t\\
&=\bm{C}\left[\bm{V},\bm{V}^*\right]\left[{\frac{\bm{g}_t^T}{2}},{\frac{\bm{g}_t^T}{2}}\right]^T,
\end{split}
\end{align}
which indicates
\begin{align}\label{eq:dtreal}
\begin{split}
\bm{d}_t=\left[\bm{V},\bm{V}^*\right]\left[{\frac{\bm{g}_t^T}{2}},{\frac{\bm{g}_t^T}{2}}\right]^T.
\end{split}
\end{align}
This means that \eqref{eq:dtreal} turns out be an alternative possible solution of \eqref{eq:DynamicANMSDP}. Accordingly, different from \eqref{eq:T=VRg2V}, the estimated $\text{T}(\bm{u}_d)$ leads to another Vandermonde decomposition as
\begin{align}\label{eq:Tudreal}
\begin{split}
\text{T}(\bm{u}_d)=\frac{1}{2}\left([\bm{V},\bm{V}^*]\;\text{diag}\left((T\bm{R}_g)^{\frac{1}{2}},(T\bm{R}_g)^{\frac{1}{2}}\right)[\bm{V},\bm{V}^*]^H\right).
\end{split}
\end{align}
Further, 
since $\bm{R}_g$ is a diagonal matrix with positive elements, substituting \eqref{eq:A=CrealV} into \eqref{eq:Rh}, we have
\begin{align}\label{eq:Rhreal}
\begin{split}
\bm{R}_h&=\bm{C}\,\text{Real}(\bm{V})\bm{R}_g\,\text{Real}(\bm{V})^H\bm{C}^H\\
&=\bm{C}\,\text{Real}(\bm{V}\bm{R}_g\bm{V}^H)\,\bm{C}^H\\
&=\bm{C}\,\text{Real}(\bm{R}_v)\,\bm{C}^H.
\end{split}
\end{align}
Noticeably, instead of \eqref{eq:Rv=VRgV}, $\text{Real}(\bm{R}_v)$ is decomposed as
\begin{align}
\begin{split}
\text{Real}(\bm{R}_v)&=\text{Real}(\bm{V}\bm{R}_g\bm{V}^H)\\
&=\frac{1}{2}\left(\bm{V}\bm{R}_g\bm{V}^H+\bm{V}^*\bm{R}_g(\bm{V}^H)^*\right)\\
&=\frac{1}{2}\left([\bm{V},\bm{V}^*]\,\text{diag}(\bm{R}_g,\bm{R}_g)[\bm{V},\bm{V}^*]^H\right).
\end{split}
\end{align}

As a result, beside the estimation of the true AoAs as $\{\hat{\theta}_k\}$, spurious results are also generated as $\{-\hat{\theta}_k\}$.
According to the above analysis that there exist multiple solutions, we need to use the prior knowledge that the feasible domain of AoAs is $[0, \pi)$. Then we can simply remove the spurious results appearing in $(-\pi, 0)$.

\subsection{Extension to the Multiple-Antenna MS Case}\label{sect:multi-MS-antenna}
In this subsection, we extend the work to the case where the MS also has a hybrid architecture with multiple RF chains and antennas in arbitrary arrays. Now, the 1D uplink channel model for the case of the single-antenna at the MS in \eqref{eq:channel} is extended to a 2D uplink channel model
\begin{equation}\label{eq:MIMOchannel}
  \bm{H}=\sum_{k=1}^K g_{k}\bm{a}_{\text{BS}}\left(\theta_{\text{BS},k}\right) \bm{a}_{\text{MS}}^H\left(\theta_{\text{MS},k}\right),
\end{equation}
where $\theta_{\text{MS},k}$ and $\theta_{\text{BS},k}$ denote the continuously-valued AoD and AoA of the $k$-th path at the MS as transmitter and at the BS as receiver, respectively. Accordingly, the vectors $\bm{a}_{\text{MS}}\left(\theta_{\text{MS},k}\right)$ and $\bm{a}_{\text{BS}}\left(\theta_{\text{BS},k}\right)$ represent the array manifold vectors corresponding to the $k$-th path for the $N_{\text{MS}}$-antenna and $N_{\text{BS}}$-antenna arrays, 
respectively, which both have 
components in the form of \eqref{eq:arbitrary}. This 2D channel model is general enough to subsume the multi-user case where each column of $\bm{H}$ corresponds to one MS (user) with a single antenna.

Given the hybrid structures and the arbitrary arrays employed at both the MS and BS sides, the received signal can be expressed as
\begin{equation}\label{eq:Y2D}
  \bm{Y} = \bm{W}\bm{A}_{\text{BS}} \, \mbox{diag}(\bm{g}) \, \bm{A}_{\text{MS}}^H \bm{F}  +  \bm{W} \bm{N},
\end{equation}
where $\bm{F}$ denotes the hybrid precoding matrix used at the transmitter side.

Applying the array manifold separation approach described in \eqref{eq:A=CV}, \eqref{eq:Y2D} can be rewritten as
\begin{align}\label{eq:Y2D-virtual}
\begin{split}
  \bm{Y} &= \bm{W}\bm{C}_{\text{BS}}\bm{V}_{\text{BS}} \, \mbox{diag}(\bm{g}) \, \bm{V}_{\text{MS}}^H \bm{C}_{\text{MS}}^H \bm{F}  +  \bm{W} \bm{N}, \\
  &= \bm{W}\bm{C}_{\text{BS}} \, \bm{\Psi} \, \bm{C}_{\text{MS}}^H \bm{F}  +  \bm{W} \bm{N},
\end{split}
\end{align}
where $\bm{\Psi} = \bm{V}_{\text{BS}} \, \mbox{diag}(\bm{g}) \, \bm{V}_{\text{MS}}^H$ denotes the virtual 2D channel that presents the 2D Vandermonde structure in the virtual uniform antenna domain.

To estimate the 2D channel through gridless CS, a straightforward way is to vectorize the 2D formulations and then to cast the 2D Vandermonde structure into a vectorized SDP formulation via a two-level Toeplitz structured matrix, a.k.a., vectorization based ANM (V-ANM) \cite{Wang2017Efficient, Chi2015Compressive, Yang2016Vandermonde}. However, the V-ANM leads to a high computational complexity on the order of $\mathcal{O}(N_{\text{BS}}^{3.5}N_{\text{MS}}^{3.5})$ \cite{Zhang2018Efficient}, 
because of the huge problem scale resulting 
from the vectorization operation.

To solve this problem of V-ANM, we develop an efficient 2D channel estimation at much lower computational cost, by using a decoupled-ANM (D-ANM) technique \cite{Tian2017Decouple, Zhang2018Efficient}. Different from the V-ANM, we introduce a matrix-form atom set $\mathcal{A}_{\text{M}}$ as
\begin{equation}
  \textstyle \mathcal{A}_{\text{M}} = \left\{ \bm{v}_{\text{BS}}(f_{\text{BS}}) \, \bm{v}_{\text{MS}}^H(f_{\text{MS}})  \left| \, f_{\text{BS}} {\in} \left( -\frac{1}{2}, \frac{1}{2}\right], f_{\text{MS}} {\in} \left( -\frac{1}{2}, \frac{1}{2}\right] \right. \right\},
\end{equation}
which naturally results in a matrix-form atomic norm as
\begin{equation}
   \|{\bm{\Psi}}\|_{\mathcal{A}_{\text{M}}} = \inf\left\{ \sum_l |g_l| \left| \, \bm{\Psi} = \sum_l g_l \, \bm{v}_{\text{BS}}(f_{\text{BS},l}) \, \bm{v}_{\text{MS}}^H(f_{\text{MS},l}) \right. \right\}.
\end{equation}
Then, given $\bm{Y}$ from \eqref{eq:Y2D-virtual}, the virtual 2D channel $\bm{\Psi}$ can be reconstructed via the following decoupled SDP formulation
\begin{align}\label{eq:DecoupledANM}
\begin{split}
  &(\hat{\bm{\Psi}}, \text{T}(\hat{\bm{u}}_{\text{BS}}), \text{T}(\hat{\bm{u}}_{\text{MS}})) =\\
&\arg\min_{\hat{\bm{\Psi}}, \text{T}(\bm{u}_{\text{BS}}), \text{T}(\bm{u}_{\text{MS}})} \frac{1}{2\sqrt{N_{\text{BS}}N_{\text{MS}}}}\left(\text{tr}(\text{T}(\bm{u}_{\text{BS}})){+}\text{tr}(\text{T}(\bm{u}_{\text{MS}})) \right) \\
&{+}\frac{\tau}{2}\normf{\bm{Y}{-}\bm{W}\bm{C}_{\text{BS}} \bm{\Psi} \bm{C}_{\text{MS}}^H \bm{F}},
~\text{s.t.}\left[\begin{array}{cc}
\text{T}(\bm{u}_{\text{BS}})&\bm{\Psi}\\
\bm{\Psi}^H&\text{T}(\bm{u}_{\text{MS}})
\end{array}\right]{\succeq} 0.\\
\end{split}
\end{align}

Noticeably, since the PSD constraint in \eqref{eq:DecoupledANM} is of size $(N_{\text{BS}}+N_{\text{MS}})\times(N_{\text{BS}}+N_{\text{MS}})$, the D-ANM allows a reduced computational complexity on the order of $\mathcal{O}\left((N_{\text{BS}}+N_{\text{MS}})^{3.5}\right)$ \cite{Zhang2018Efficient}, which is much smaller than that of the V-ANM with large arrays.

\section{Numerical Results}\label{sect:simulations}
This section presents numerical results to evaluate the channel estimation performance achieved by the proposed methods for arbitrary arrays and a hybrid precoding structure. In each Monte Carlo simulation, the random path angles are generated uniformly from $[0^{\circ}, 180^{\circ})$. The existing channel covariance estimation methods via covariance orthogonal matching pursuit (COMP) and Dynamic-COMP (DCOMP) and the existing instantaneous CSI estimation methods via simultaneous orthogonal matching pursuit (SOMP) and Dynamic-SOMP (DSOMP) are also simulated as benchmarks for performance comparison \cite{Park2018Spatial}, where a predefined grid of size $360$ is employed for the grid-based CS technique that leads to an 
angle-resolution of $0.5^{\circ}$.

\subsection{Channel Estimation Performance}
First, we testify the performance of different channel estimation approaches with an arbitrary planar array, in terms of the normalized mean squared error (NMSE) for the CCE as $\mathbb{E}\{{\|\bm{R}_h-\hat{\bm{R}}_h\|_F^2}\}/\mathbb{E}\{{\|\bm{R}_h\|_F^2}\}$ and the NMSE for the CSI estimation as $\mathbb{E}\{{\|\bm{h}-\hat{\bm{h}}\|_2^2}\}/\mathbb{E}\{{\|\bm{h}\|_2^2}\}$, respectively. In simulations, our LRSCR-based methods as described in Section~\ref{sect:LRSCR} and the COMP-based and SOMP-based methods in \cite{Park2018Spatial}
are tested on the fixed combining matrix $\bm{W}$, while our Dynamic-ANM-based methods as developed in Section~\ref{sect:DynamicANM} and the DCOMP-based and the DSOMP-based methods in \cite{Park2018Spatial} are applied with the dynamic combining matrix $\bm{W}_t$. In our proposed two-step CSI estimation scheme, we employ a Vandermonde decomposition in the form of either \eqref{eq:T=VRg2V} or \eqref{eq:Rv=VRgV} to retrieve the path angles in the first step and to estimate the path gains via least squares in \eqref{eq:LS} in the second step.

\begin{figure}
	\centering \includegraphics[width=3.5in]{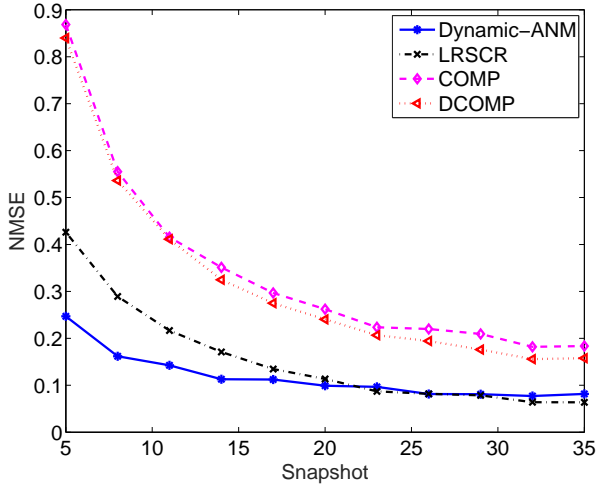}
		\caption{NMSE of CCE versus $T$ for LRSCR, Dynamic-ANM, COMP and DCOMP, when $N{=}64, M{=}16, I{=}35, K{=}4, \text{SNR}{=}10$dB.}
		\label{fig:CCE_vs_snapshot}
\end{figure}
\begin{figure}
	\centering \includegraphics[width=3.5in]{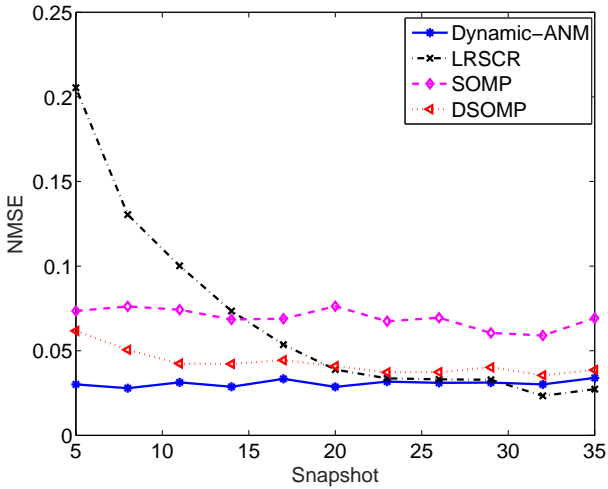}
		\caption{NMSE of CSI estimation versus $T$ for LRSCR, Dynamic-ANM, SOMP and DSOMP, when $N{=}64, M{=}16, I{=}35, K{=}4, \text{SNR}{=}10$dB.}
		\label{fig:CSIE_vs_snapshot}
\end{figure}
\begin{figure}
	\centering \includegraphics[width=3.5in]{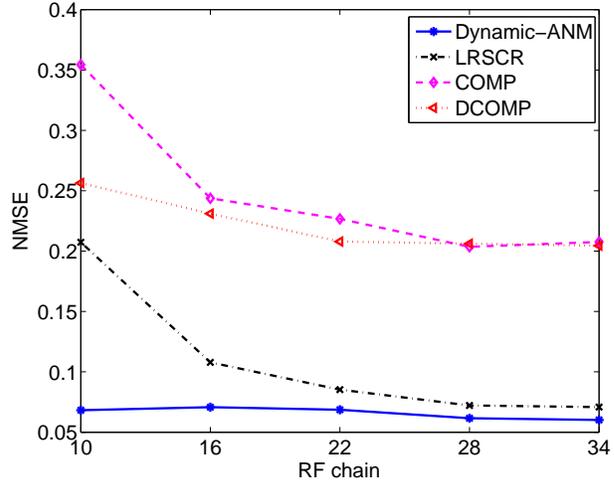}
		\caption{NMSE of CCE versus $M$ for LRSCR, Dynamic-ANM, COMP and DCOMP, when $N{=}64, T{=}20, I{=}35, K{=}4, \text{SNR}{=}10$dB.}
		\label{fig:CCE_vs_RFchain}
\end{figure}
\begin{figure}
	\centering \includegraphics[width=3.5in]{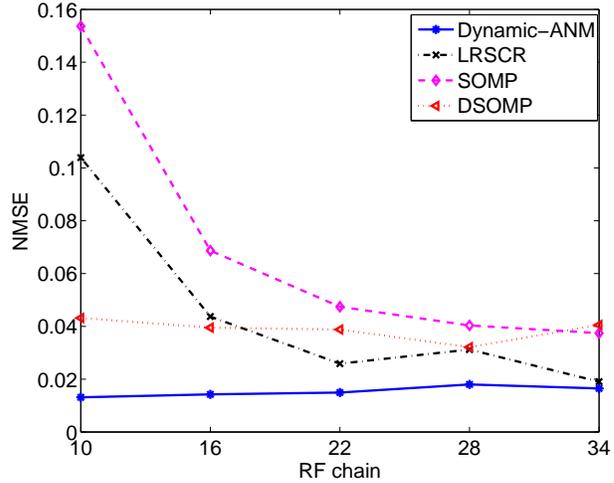}
		\caption{NMSE of CSI estimation versus $M$ for LRSCR, Dynamic-ANM, SOMP and DSOMP, when $N{=}64, T{=}20, I{=}35, K{=}4, \text{SNR}{=}10$dB.}
		\label{fig:CSIE_vs_RFchain}
\end{figure}

Fig.~\ref{fig:CCE_vs_snapshot} and Fig.~\ref{fig:CSIE_vs_snapshot} present the NMSE of CCE and the NMSE of CSI versus the number of snapshots, respectively. The comparison of the curves indicates that our proposed methods based on LRSCR and Dynamic-ANM 
outperform the existing methods based on 
grid-based CS.
While the dynamic configurations usually provide a higher sensing accuracy than the fixed counterparts, our LRSCR can even work better than 
DSOMP as the number of snapshots increases.
This is because our methods utilize not only the sparsity of mmWave channels but also the structural feature of the array geometry. Further, our Dynamic-ANM method always achieves the best performance especially given a small number of snapshots, because ANM can efficiently utilize such structures directly from the collected samples. On the other hand, since LRSCR is a statistics-based design, it requires a sufficient number of snapshots for computing an accurate sample covariance. When the number of snapshots becomes small e.g. less than $20$, the finite-sample effect ruins the Toeplitz structure presented in the ideal covariance matrix, which thus degrades the performance of LRSCR.
Fig.~\ref{fig:CCE_vs_RFchain} and Fig.~\ref{fig:CSIE_vs_RFchain} present the NMSE performance of these approaches for different numbers of RF chains, which show 
the same trends as in Fig.~\ref{fig:CCE_vs_snapshot} and Fig.~\ref{fig:CSIE_vs_snapshot}.

\subsection{Nonuniform Linear Array} 
\begin{figure}
	\centering \includegraphics[width=3.5in]{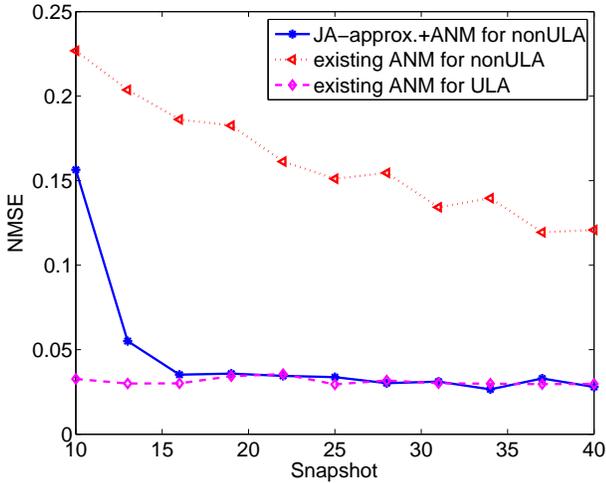}
		\caption{NMSE of CSI estimation versus $T$ for Dynamic-ANM, existing ANM for nonuniform linear array, and existing ANM for the same aperture size ideal uniform linear array, when $N{=}16, M{=}8, I{=}60, K{=}4, \text{SNR}{=}10$dB.}
		\label{fig:JAapprox}
\end{figure}

Next, we study the nonuniform linear array as a special case of arbitrary arrays. Fig.~\ref{fig:JAapprox} shows that the NMSE of the instantaneous CSI estimation via our Dynamic-ANM method can be much smaller 
than that of the existing ANM without array manifold separation \cite{Wang2017Efficient}, which demonstrates the necessity of the Jacobi-Anger approximation for an imperfect array geometry.
Besides, it also indicates that our method proposed for arbitrary arrays can approach the performance of the ideal case of a same-size ULA as the benchmark for the best performance that the proposed techniques can achieve.

\begin{figure}
	\centering
		\includegraphics[width=3.5in]{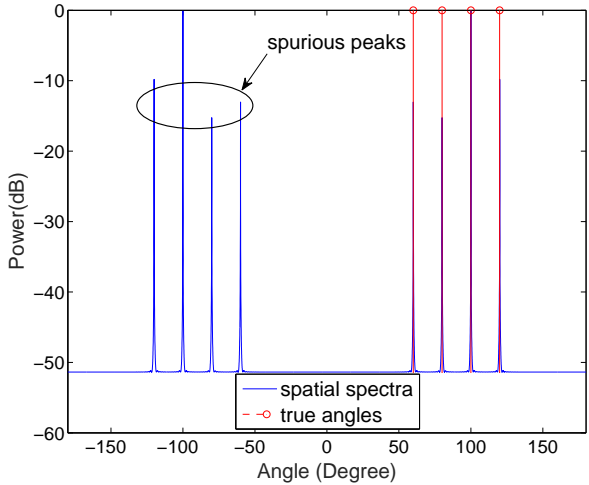}
		\caption{Spatial spectra for LRSCR with nonuniform linear array}
		\label{fig:spurious-peak}
\end{figure}

Further, 
we study the side effect of the Jacobi-Anger approximation in terms of the spurious peaks generated in nonuniform linear arrays. Fig.~\ref{fig:spurious-peak} shows the spatial spectra result of LRSCR, where the peaks on the right indicate the true angles. Meanwhile, the spurious peaks appear at the symmetric angles, which can be simply removed given the prior knowledge of the angle range as $[0^{\circ}, 180^{\circ})$.

\subsection{Computational Complexity}
\begin{figure}
	\centering \includegraphics[width=3.5in]{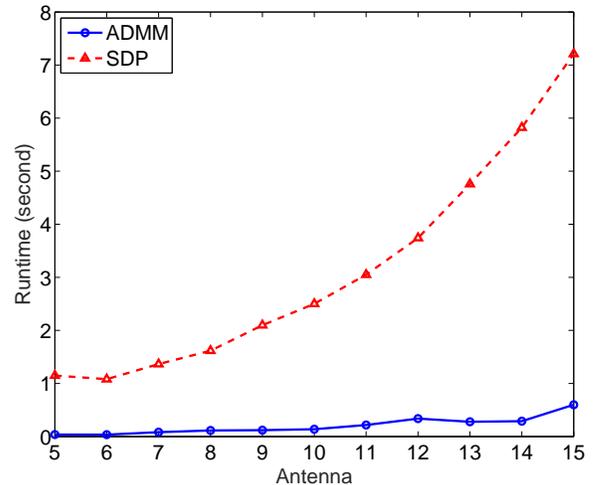}
        \caption{Runtime versus $N$ for ADMM and SDP implementations.}
		\label{fig:runtime}
\end{figure}

In addition, we test the computational cost of the proposed fast algorithm via ADMM, compared with the SDP-based solver \cite{Grant2018CVX}. By counting the runtime versus the number of antennas, Fig.~\ref{fig:runtime} clearly indicates that as the number of antennas increases the slope of the runtime curve of the ADMM-based solution is much smaller than that of the SDP counterpart. Thus, the proposed fast implementation has low computational complexity and is well suited for large arrays.

\subsection{2D Path Angle Estimation} 
\begin{figure}
	\centering \includegraphics[width=3.5in]{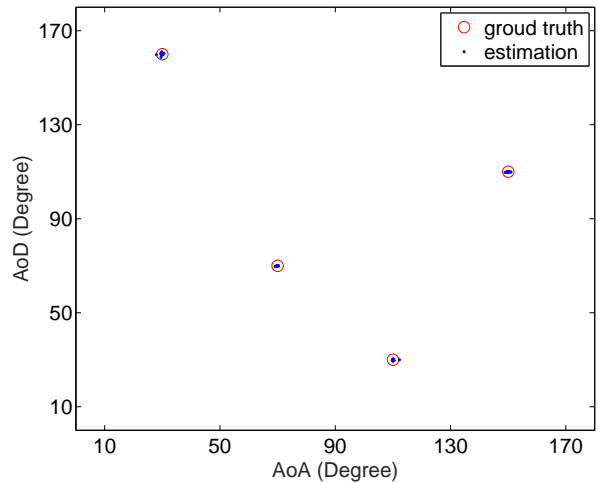}
		\caption{2D path angle estimation via D-ANM in a multiple-antenna MS case with $N_{\text{MS}}{=}32$, $N_{\text{BS}}{=}32$ and $\text{SNR}{=}10$dB.}
		\label{fig:2D-AngleEst}
\end{figure}

Last but not least, we extend our proposed work to the 
multiple-antenna 
MS case
where both the BS and MS are equipped with arbitrary multiple antennas, which results in a 2D path angle estimation scenario. As shown in Fig.~\ref{fig:2D-AngleEst}, our proposed D-ANM based on the Jacobi-Anger approximation can precisely retrieve both the AoAs and the AoDs of the sparse scattering paths, which indicates the high performance of our proposed super-resolution 2D channel estimation techniques.

\section{Conclusions}\label{sect:conclusions}
Recognizing the imperfect geometry of arbitrary arrays and the hardware constraint of hybrid mmWave massive MIMO systems, this paper has proposed a new super-resolution channel estimation framework that achieves the benefits of the array manifold separation techniques and the structure-based optimization approaches. Through the Jacobi-Anger approximation, the Vandermonde structure is effected in the virtual antenna domain of arbitrary arrays, which enables super-resolution channel estimation based on gridless CS techniques to obtain a high performance at low training costs. In particular, we develop two channel covariance estimation approaches via LRSCR and Dynamic-ANM. Further, considering that angles change relatively slower than 
path gains, we design a two-step CSI estimation scheme which separates long-term angle estimation from frequent path gain estimation. The theoretical results are provided to investigate the fundamental limits of the proposed super-solution technique in terms of the minimum number of RF chains required for channel estimation and the bounds on the mode order for the Jacobi-Anger approximation. To reduce the computational complexity of the structure-based optimization via SDP, a first-order iterative algorithm is developed through ADMM for fast implementation. To combat the side effect due to the Jacobi-Anger approximation occurring in 
nonuniform linear arrays, we provide a mechanism to efficiently remove the spurious peaks. Finally, we extend our work to the 2D-angle scenarios, where both the BS and MS are equipped with multiple antennas.



\end{document}